\input harvmac 
\input epsf.tex

\overfullrule=0mm

\newcount\figno
\figno=0
\def\fig#1#2#3{
\par\begingroup\parindent=0pt\leftskip=1cm\rightskip=1cm\parindent=0pt
\baselineskip=11pt
\global\advance\figno by 1
\midinsert
\epsfxsize=#3
\centerline{\epsfbox{#2}}
\vskip 12pt
{\bf Fig.\the\figno:} #1\par
\endinsert\endgroup\par
}
\def\figlabel#1{\xdef#1{\the\figno}}
\def\encadremath#1{\vbox{\hrule\hbox{\vrule\kern8pt\vbox{\kern8pt
\hbox{$\displaystyle #1$}\kern8pt}
\kern8pt\vrule}\hrule}}


\def\IR{\relax{\rm I\kern-.18em R}}
\font\cmss=cmss10 \font\cmsss=cmss10 at 7pt

\font\cmss=cmss10 \font\cmsss=cmss10 at 7pt
\def\IZ{\relax\ifmmode\mathchoice
{\hbox{\cmss Z\kern-.4em Z}}{\hbox{\cmss Z\kern-.4em Z}}
{\lower.9pt\hbox{\cmsss Z\kern-.4em Z}}
{\lower1.2pt\hbox{\cmsss Z\kern-.4em Z}}\else{\cmss Z\kern-.4em Z}\fi}
\def\IN{\relax{\rm I\kern-.18em N}}
\def\b{\circ}
\def\n{\bullet}

\def\gbbbb{\Gamma_4^{\hbox{$\scriptstyle \b \b$}\kern -8.2pt
\raise -4pt \hbox{$\scriptstyle \b \b$}}}
\def\gnnnn{\Gamma_4^{\hbox{$\scriptstyle \n \n$}\kern -8.2pt  
\raise -4pt \hbox{$\scriptstyle \n \n$}}}
\def\gnnnnnn{\Gamma_6^{\hbox{$\scriptstyle \n \n \n$}\kern -12.3pt
\raise -4pt \hbox{$\scriptstyle \n \n \n$}}}
\def\gbbbbbb{\Gamma_6^{\hbox{$\scriptstyle \b \b \b$}\kern -12.3pt
\raise -4pt \hbox{$\scriptstyle \b \b \b$}}}
\def\gbbbbc{\Gamma_{4\, c}^{\hbox{$\scriptstyle \b \b$}\kern -8.2pt
\raise -4pt \hbox{$\scriptstyle \b \b$}}}
\def\gnnnnc{\Gamma_{4\, c}^{\hbox{$\scriptstyle \n \n$}\kern -8.2pt
\raise -4pt \hbox{$\scriptstyle \n \n$}}}
\def\Rbud#1{{\cal R}_{#1}^{-\kern-1.5pt\blacktriangleright}}
\def\Rleaf#1{{\cal R}_{#1}^{-\kern-1.5pt\vartriangleright}}
\def\Rbudb#1{{\cal R}_{#1}^{\circ\kern-1.5pt-\kern-1.5pt\blacktriangleright}}
\def\Rleafb#1{{\cal R}_{#1}^{\circ\kern-1.5pt-\kern-1.5pt\vartriangleright}}
\def\Rbudn#1{{\cal R}_{#1}^{\bullet\kern-1.5pt-\kern-1.5pt\blacktriangleright}}
\def\Rleafn#1{{\cal R}_{#1}^{\bullet\kern-1.5pt-\kern-1.5pt\vartriangleright}}
\def\Wleaf#1{{\cal W}_{#1}^{-\kern-1.5pt\vartriangleright}}
\def\Cleaf{{\cal C}^{-\kern-1.5pt\vartriangleright}}
\def\Cbud{{\cal C}^{-\kern-1.5pt\blacktriangleright}}
\def\Crleaf{{\cal C}^{-\kern-1.5pt\circledR}}


\Title{\vbox{\hsize=3.truecm \hbox{SPhT/03-086}}}
{{\vbox {
\bigskip
\centerline{Random trees between two walls:}
\medskip
\centerline{Exact partition function} 
}}}
\bigskip
\centerline{J. Bouttier\foot{bouttier@spht.saclay.cea.fr}, 
P. Di Francesco\foot{philippe@spht.saclay.cea.fr} and
E. Guitter\foot{guitter@spht.saclay.cea.fr}}
\medskip
\centerline{ \it Service de Physique Th\'eorique, CEA/DSM/SPhT}
\centerline{ \it Unit\'e de recherche associ\'ee au CNRS}
\centerline{ \it CEA/Saclay}
\centerline{ \it 91191 Gif sur Yvette Cedex, France}
\bigskip
\noindent 
We derive the exact partition function for a discrete model of random
trees embedded in a one-dimensional space.
These trees have vertices labeled by integers representing their 
position in the target space, with the SOS constraint that 
adjacent vertices have labels differing by $\pm 1$.
A non-trivial partition function is obtained whenever the target space
is bounded by walls. We concentrate on the two cases where
the target space is (i) the half-line bounded by 
a wall at the origin or (ii) a segment bounded by two walls at
a finite distance.  
The general solution has a soliton-like structure involving 
elliptic functions. We derive the corresponding continuum scaling limit 
which takes the  remarkable form of the Weierstrass $\wp$ function
with constrained periods. These results are used to analyze the 
probability for an evolving population spreading in one dimension 
to attain the boundary of a given domain with the geometry of
the targets (i) or (ii).
They also translate, via suitable bijections, 
into generating functions for bounded planar graphs.

\Date{06/03}

\nref\LEGAL{See e.g. J.-F. Le Gall,{\it Spatial Branching Processes, Random Snakes and Partial
Differential Equations}, Birkhauser, Boston (1999).}
\nref\CS{P. Chassaing and G. Schaeffer, {\it Random Planar Lattices and 
Integrated SuperBrownian Excursion}, preprint (2002), to appear in 
Probability Theory and Related Fields, math.CO/0205226.}
\nref\GEOD{J. Bouttier, P. Di Francesco and E. Guitter, {\it Geodesic
distance in planar graphs}, Nucl. Phys. {\bf B663} [FS] (2003) 535-567.}
\nref\JM{See e.g. M. Jimbo and T. Miwa, {\it Solitons and infinite dimensional Lie
algebras}, Publ. RIMS, Kyoto Univ. {\bf 19} No. 3 (1983) 943-1001, 
eq.(2.12).}
\nref\BAT{H. Bateman, {\it Higher Transcendental Functions}, vol. II,
McGraw-Hill, New-York (1953).}
\nref\GALWA{S. Karlin and H. Taylor, {\it A first course in
stochastic processes}, Academic Press, New-York (1975).}
\nref\SCH{G. Schaeffer, {\it Bijective census and random 
generation of Eulerian planar maps}, Electronic
Journal of Combinatorics, vol. {\bf 4} (1997) R20; see also
G. Schaeffer, {\it Conjugaison d'arbres
et cartes combinatoires al\'eatoires} PhD Thesis, Universit\'e 
Bordeaux I (1998).}
\nref\BMS{M. Bousquet-M\'elou and G. Schaeffer,
{\it Enumeration of planar constellations}, Adv. in Applied Math.,
{\bf 24} (2000) 337-368; see also D. Poulalhon and G. Schaeffer,
{\it A note on bipartite Eulerian planar maps}, preprint (2002), available 
at {\sl http://www.loria.fr/$\sim$schaeffe/} and
J. Bouttier, P. Di Francesco and E. Guitter,
{\it Counting colored Random Triangulations},
Nucl.Phys. {\bf B641} (2002) 519-532.}
\nref\LERETOUR{J. Bouttier, P. Di Francesco and E. Guitter, 
{\it Statistics of planar graphs viewed from a vertex: A study via labeled trees},
preprint cond-mat/0307606 and SPhT/03-104 (2003), to appear in Nucl. 
Phys. B.} 
\nref\DELMAS{J.-F. Delmas {\it Computation of moments for the length of the
one dimensional ISE support}, preprint (2002), available at
http://cermics.enpc.fr/$\sim$delmas/.}

\newsec{Introduction}

\fig{A sample rooted labeled planar tree, with root (top) vertex labeled by $2$.
Neighboring vertices have labels differing by $\pm 1$. These labels
may be viewed as positions on a target integer line as indicated.}{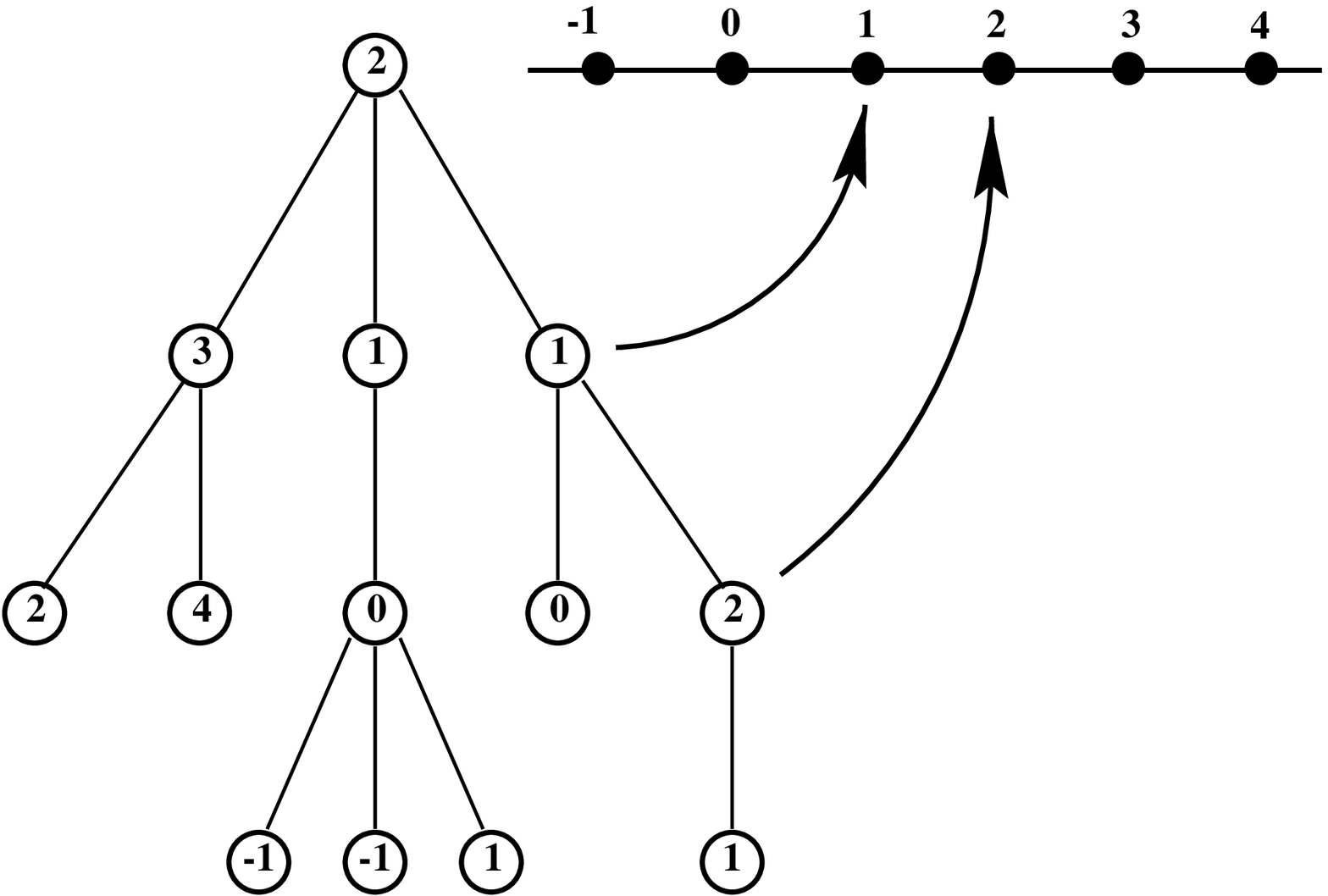}{11.cm}
\figlabel\labtree

In this paper, we consider a simple model describing the embedding 
in one dimension of a random tree. More precisely, we consider
random {\it rooted} trees whose vertices are labeled by
integers representing their possible discrete positions in a 
one-dimensional target space. We moreover impose that two neighboring 
vertices on the tree have labels differing by $+1$ or $-1$, which allows
to view the edges of the tree as rigid segments of unit length embedded
in the real line (see Fig.\labtree\ for an illustration). 
We choose to consider the case of so-called
{\it planar} trees, i.e. we count as {\it distinct} all trees obtained by
permuting any two descendent subtrees at a given vertex. 
This model is nothing but a discrete version
of the so called one-dimensional Brownian snake \LEGAL\ which
is used to describe branching processes, for instance the spreading of a population
in a one-dimensional target space. 
In this language, we may think of our
rooted trees as representing ``genealogical trees'' for the
lineage of an initial individual (materialized by the root vertex), 
while the labels represent
the positions in space of all the descendent individuals. The spreading 
process is modeled here by demanding that each individual lives at distance
one from its parent\foot{The particular discrete spreading rule chosen here 
should not affect the universality of the continuum answer.}. 
We will discuss this interpretation in detail
in section 5 below. 

Alternatively, we may view this model as a statistical 
Solid-On-Solid (SOS) model with heights given by the labels, and whose 
base space is a random tree. The SOS rule of having neighboring heights 
differing by $\pm 1$ is responsible for the ``integrability" of the model.
However, as opposed to the two-dimensional lattice case where a roughening transition
takes place, we expect here that the discrete nature of the heights is eventually 
irrelevant when working with large tree-like base spaces. 
 

We shall consider a statistical 
sum over all such trees with a weight $g$ per edge, and with a fixed 
position, say $n$, of the root vertex. With no bounds on the labels, 
the partition function is trivial as it amounts 
to counting rooted planar trees with $N$ edges (in number $c_N$ where
$c_{N}={2N\choose N}/(N+1)$ is the $N$-th Catalan number), 
each of which gives rise to $2^{N}$ possible embeddings. Similarly, the
width $w_N=\sqrt{\langle n^2\rangle_N}$ for the fluctuations of the labels
$n$ in a random tree of size $N$ and with root label $0$ is easily obtained
as
\eqn\wnres{ w_N^2={1\over 2}\left({4^N\over {2N \choose N}}-1\right) }
hence $w_N\sim (\pi N/4)^{1/4}$ for large $N$. 

The problem becomes more interesting in the presence of a wall, say at 
position $-1$, which amounts to imposing that all labels be non-negative.
Similar trees were introduced under the name of ``well-labeled trees''
in Ref.\CS\ in connection with the enumeration of rooted planar tetravalent
graphs, with vertex labels representing the (necessarily non-negative)
geodesic distance on the graph to the root vertex\foot{In this reference, 
a slightly different constraint is imposed on the labels, demanding that 
neighboring vertices have labels differing by $0,\pm 1$. Our results will
be easily extended to this modified case in Sect.6 below, with no
fundamental difference.}. The explicit
form of the corresponding partition function as a function of $g$ and
$n$ was given in Ref.\GEOD\ in the context of planar graph enumeration and will
be recalled in the next section. A remarkable outcome of this solution
is the emergence of discrete soliton-like expressions, suggesting an
underlying integrable structure.

The purpose of this paper is to study the statistics of trees
with labels now belonging to a {\it finite set}, say 
$\{0,1,\cdots,L\}$, which amounts to having two walls at positions 
$-1$ and $L+1$ in the embedding target space.
In the language of evolving populations, 
introducing walls gives access to the probability for the population to
attain pre-defined boundaries or to remain confined within a pre-defined 
connected domain. The two-wall situation corresponds
to the generic case where this domain is compact, i.e. is a segment (with two boundaries).    
Such boundary conditions correspond to the so-called 
Restricted Solid-On-Solid (RSOS) version of the problem, in which 
heights are restricted to belong to a finite segment.

The paper is organized as follows. 
In Sect.2, we introduce a master equation for the partition function of 
labeled rooted trees and recall its solutions both without walls
and with one wall. Sect.3 is devoted to the derivation of the two-wall
solution, involving elliptic functions. The corresponding
continuum limit is derived in Sect.4 and expressed in terms of
the Weierstrass $\wp$ function with constrained periods. 
As an application of these results
we study in Sect.5 a particular stochastic process
describing the evolution of a population which spreads in one dimension. 
We discuss in Sect.6
the solution of a slightly different problem
corresponding to a dilute SOS version in which
neighboring vertices of the tree have labels differing by
$\pm 1$ {\it or} $0$.
We gather a few concluding remarks in Sect.7, where we discuss the integrability of
our models in particular in connection with planar graph enumeration and
matrix models. 
The precise connection between labeled trees and planar graphs is
further detailed in Appendix A.

\newsec{Enumeration of labeled rooted trees}

\fig{A sketch of the master equation (2.1). A tree contributing to $R_n$
is decomposed according to the sequence of descendents of its root
(labeled by $n$). Each descendent vertex is 
arbitrarily labeled by $n\pm 1$ henceforth the descendent
subtrees are generated by $R_{n\pm 1}$ accordingly. Each edge connected to the root 
is given a weight $g$. 
The summation over all possible configurations produces the r.h.s.
of eq.(2.1).}{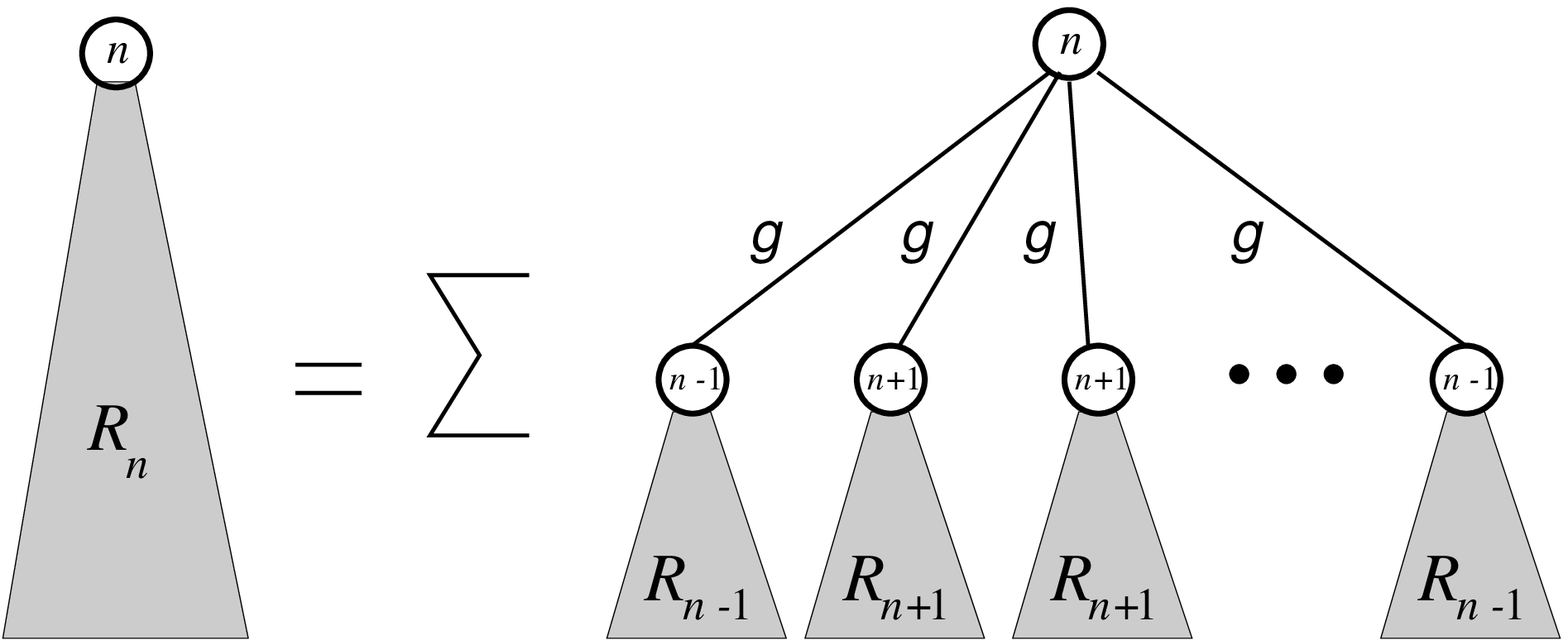}{12.cm}
\figlabel\environ

Let $R_n$ denote the generating function for labeled rooted
planar trees with a root at position $n$. By a decomposition according to the 
possible local environments of the root vertex, characterized
by the sequence of labels $n-1$ or $n+1$ of its adjacent vertices (see Fig.\environ), 
we immediately get the equation
\eqn\recu{R_n={1\over 1-g(R_{n+1}+R_{n-1})}}
Note also that from their combinatorial definition as counting functions,
the $R_n$'s are required to have a series expansion in powers of $g$ starting
as $R_n=1+O(g)$, a condition sufficient to determine all the $R_n$'s from
eq.\recu. The relation \recu\ is valid for all accessible values
of $n$ and it may be supplemented by boundary conditions to account
for the possible presence of walls. We choose those conditions among 
three categories: no wall, one wall and two walls,
corresponding to restrictions on the allowed labels $n$, respectively
no restriction, $n\geq 0$ and $0\leq n\leq L$.

In the absence of wall, all the $R_n$'s are equal due to translational
invariance, to a function $R$ satisfying $R=1/(1-2gR)$ and $R=1+O(g)$,
namely:
\eqn\cata{R=R(g)\equiv {1-\sqrt{1-8g}\over 4g}=\sum_{N=0}^{\infty}g^N 2^N c_{N}}
This formula displays the critical value $g_c=1/8$ of $g$, while the coefficient
of $g^N$ clearly counts rooted planar trees with $N$ edges ($c_N$) with $2^N$
possible embeddings.

In the presence of one wall at position $-1$, we must write $R_{-1}=0$
and consider the equation \recu\ only for $n\geq 0$.
This system may be solved order by order in $g$ using
as a seed the vanishing of all the coefficients of the series
for $R_{-1}$ and the order zero values $R_n=1+O(g)$ for all $n\geq 0$.
In a more global way, the solution was worked out in Ref.\GEOD\ by 
replacing the condition of existence of a power series expansion for 
each $R_n$ by the condition that $R_n\to R$ at large $n$ with $R$ 
as above. Indeed, it is clear from the combinatorial definition
that $R-R_n=O(g^{n+1})$ as the wall at position $-1$ may not be reached 
with less than $(n+1)$ edges from position $n$. The solution of
eq.\recu\ with these boundary conditions reads:
\eqn\onewsolu{R_n = R {u_n u_{n+4} \over u_{n+1} u_{n+3}}, \quad
u_n=x^{n+1\over 2}-x^{-{n+1\over 2}}}
with $R=R(g)$ as in eq.\cata\ and where $x$ is the solution of
\eqn\chara{x+{1\over x}+2={1\over g R}}
with, say, modulus less than $1$, namely 
$x=x(g)\equiv (1-(1-8g)^{1\over 4})/(1+(1-8g)^{1\over 4})$. Note that $x$
is real for all $g\leq 1/8$ and admits a convergent series expansion 
in $g$ with {\it positive integer} coefficients. The small $g$ behavior
$x(g)=g+O(g^2)$ ensures the above property that $R_n=R+O(g^{n+1})$. 
The solution \onewsolu\ is readily checked by noting that eq.\recu\
translates into the following trilinear equation for the $u$'s 
\eqn\trilin{u_n u_{n+2} u_{n+4} = {1\over R} u_{n+1}u_{n+2}u_{n+3}+g R (u_{n-1}
u_{n+3}u_{n+4}+u_nu_{n+1}u_{n+5})}
easily verified upon substituting $gR=x/(1+x)^2$ and $1/R=(1+x^2)/(1+x)^2$.
The particular form of the solution \onewsolu\ was identified in Ref.\GEOD\
as a stationary one-soliton solution to the KP equation \JM.

In the presence of two walls, we must write $R_{-1}=R_{L+1}=0$
and consider eq. \recu\ for $0\leq n\leq L$. This system 
may again be solved order by order in $g$ from the boundary conditions
that all coefficients of $R_{-1}$ and $R_{L+1}$ vanish and $R_n=1+O(g)$
for all $n$, $0\leq n\leq L$. For a finite value of $L$, one may
also eliminate all but one $R_n$ from the finite set of algebraic equations
\recu\ so as to obtain an algebraic equation of degree $[(L+3)/2]$
for each $R_n$, with a unique solution such that $R_n=1+O(g)$.
In a more global way, we intend to generalize the solution \onewsolu\
to this two-wall case. This is done in the next section.

\newsec{Two-wall solution}

\subsec{General solution via elliptic functions}

From the one-soliton structure of the solution \onewsolu, it is natural
to look for a more general soliton-like solution to the equation \recu\
in the ``elliptic'' form
\eqn\twowsolu{\eqalign{
R_n &= R {u_n u_{n+4} \over u_{n+1} u_{n+3}} \cr
u_n&= (x^{n+1\over 2}-x^{-{n+1\over 2}})
\prod_{j=1}^\infty (1-q^j x^{n+1})(1-{q^j\over x^{n+1}}) \cr}}
with an additional free real parameter $q$ such that $|q|<1$.
For $q=0$, we recover the solution \onewsolu\ provided $R$ and $x$ 
are given by eqns.\cata\ and \chara, depending on $g$ only. 
As we shall now see, for a general $q$, we may tune $R$ and $x$ 
in eq.\twowsolu\ as functions of {\it both} $g$ {\it and} $q$ so as 
to satisfy the equation \recu. The above 
solution clearly satisfies $R_{-1}$=0 and the value of the free parameter
$q$ may finally be adjusted so as to ensure $R_{L+1}=0$. 

To fix the functions $R$ and $x$ in terms of $g$ and $q$, we again
write eq.\recu\ in the form of eq.\trilin, and note that from
the definition of $u_n$, we have $u_{-1}=0$ and $u_{-2-k}=-u_k$.
Hence, taking $n=-1$ in eq.\trilin, we get $g R^2= u_1/u_3$
while taking $n=-2$, we obtain $g R = (u_0/u_1)^2$, which
generalizes eq.\chara.
This leads to:
\eqn\eqnforg{g={u_0^4 u_3 \over u_1^5}}
which implicitly determines $x(g,q)$ as a function 
of $g$ and $q$. More precisely, for a fixed $q$ with $0\leq q<1$, 
this equation has four solutions for $0\leq g \leq 1/8$,
a real positive one $x_0(g,q)$ with, say, modulus less than $1$ together with its
inverse $x_2(g,q)=1/x_0(g,q)$, and a solution on the unit circle
$x_1(g,q)$ with, say, positive imaginary part together with its
inverse $x_3(g,q)=1/x_1(g,q)$. For a fixed $q$ with $-1<q\leq 0$,
we have the same pattern of solutions provided $g\geq 1/8$ and does
not exceed some upper bound.
When $g=1/8$, all solutions coalesce to $x=1$ independently of $q$.
For a definite value of the position $L$ of the second wall, 
the proper choice of solution will be discussed below. 
In particular, the presence of two walls at a finite distance increases 
the radius of convergence of the $R_n$'s as series of $g$ as it reduces 
the entropy of configurations. This requires exploring values of $g>1/8$.
Note that for $q=0$, we have $x_0(g,0)=x(g)$, corresponding to the solution
of section 2, while $x_1(g,0)=(1+i(1-8g)^{1\over 4})/(1-i(1-8g)^{1\over 4})$.

The function $R$ is now given by 
\eqn\eqnforR{R={u_1^3\over u_0^2 u_3}}
with two determinations $R^{(0)}(g,q)$ and $R^{(1)}(g,q)$ corresponding 
respectively to the substitutions $x=x_0(g,q)$ and $x=x_1(g,q)$
in the $u_n$'s, or equivalently $x_2(g,q)$ and $x_3(g,q)$ respectively
as the change $x\to 1/x$ amounts to $u_n\to -u_n$, leaving $R$ and all 
$R_n$'s invariant.
It remains to verify that, for these particular choices of $x$ and $R$, the
$u_n$'s of eq.\twowsolu\ actually satisfy the identity \trilin\ for
all $n$.
Let us introduce the notations $q=e^{2i\pi \tau}$, $x^{n+1}=e^{2i\pi z}$
so that $u_n=\theta_1(z)$ where $\theta_1$ is the (unnormalized)
Jacobi theta function with nome $q$ and argument $z$:
\eqn\thetaone{\theta_1(z)\equiv\theta(z|\tau)=2i \sin(\pi z)\prod_{j\geq 1}
(1-2 q^j \cos(2\pi z)+q^{2 j})}
We also introduce the notation $x=e^{2i\pi \alpha}$, so that eq.\trilin,
together with the particular choices \eqnforg\ and \eqnforR, translates into a
theta function identity 
\eqn\tetid{\eqalign{ \left({\theta_1(2\alpha)\over\theta_1(\alpha)}\right)^2
\theta_1(z)\theta_1(z+2\alpha)&\theta_1(z+4\alpha)
=
{\theta_1(4\alpha)\over \theta_1(2\alpha)} \theta_1(z+\alpha)\theta_1(z+2\alpha)
\theta_1(z+3\alpha)\cr
&+\theta_1(z-\alpha)\theta_1(z+3\alpha)\theta_1(z+4\alpha)
+\theta_1(z)\theta_1(z+\alpha)\theta_1(z+5\alpha)\cr}}
To prove it, we note that both hands have the same
transformations when $z\to z+1$ and $z\to z+\tau$, due to
the properties $\theta_1(z+1)=-\theta_1(z)$ and 
$\theta_1(z+\tau)= -q^{-{1\over 2}}e^{-2i\pi z} \theta_1(z)$.
Taking the ratio of rhs/lhs of eq.\tetid, we get an elliptic function, 
with poles possibly at $z=0,-2\alpha,-4\alpha$ in a fundamental cell. 
One easily checks by examining the rhs at these values that the 
corresponding residues all vanish and therefore the ratio is a constant,
easily shown to be $1$ by taking for instance its value at
$z=-\alpha$, which completes the proof of the identity. 
With these notations, we have the relations
\eqn\gforeq{ g={\theta_1^4(\alpha) \theta_1(4\alpha)\over \theta_1^5(2\alpha)}}
and
\eqn\getR{ R= {\theta_1(2\alpha)^3 \over \theta_1(\alpha)^2 \theta_1(4\alpha)}
}
while
\eqn\getRn{ R_n= {\theta_1(2\alpha)^3 \over \theta_1(\alpha)^2\theta_1(4\alpha)}
\,{\theta_1((n+1)\alpha)\theta_1((n+5)\alpha)\over\theta_1((n+2)\alpha)
\theta_1((n+4)\alpha)}}
with $\theta_1(z)$ as in eq.\thetaone.
As this solution involves elliptic functions, it is natural to
study its transformation under the modular transformation
$\tau\to -1/\tau$. From the transformation 
$\theta_1(z/\tau|-1/\tau)=-i (-i\tau)^{1/2} e^{i\pi z^2/\tau}\theta_1(z|\tau)$, we find
that the {\it physical quantities} such as $g$ as given by eq.\gforeq\ above 
and $R_n$ as given by eq.\getRn\ are invariant under 
$(\alpha,\tau)\to ({\tilde \alpha},{\tilde \tau})\equiv (\alpha/\tau,-1/\tau)$
or equivalently $(x,q)\to ({\tilde x},{\tilde q})$ with 
${\tilde x}=x^{1/\tau}$ and ${\tilde q}=e^{-2i\pi/\tau}$. 
This is not the case for intermediate factors such as $R$ or 
any of the $u_n$'s taken independently.
We deduce from these properties that the {\it same} solution is
reached by the parameters $(x,q)$ and by their modular transforms
$({\tilde x},{\tilde q})$.

It is interesting to study the modular transformation of the solutions
$x_i(g,q)$, $i=0,1,2,3$ of eq.\gforeq\ as introduced above. Clearly, 
for a fixed $g$, the modular invariance of the r.h.s. of eq.\gforeq\ 
implies that ${\tilde x_i(g,q)}=x_i(g,q)^{1/\tau}$ is a valid solution 
when $q\to {\tilde q}$, hence we may write 
$x_i(g,q)^{1/\tau}=x_{\sigma(i)}(g,{\tilde q})$ for some permutation
$\sigma\in S_4$. Iterating this transformation, and using ${\tilde {\tilde {q}}}=q$
while $\tau {\tilde \tau}=-1$, we deduce that $x_{\sigma^2(i)}(g,q)=x_i(g,q)^{-1}$,
which brings us back to the same solution we started from, but with the
determination $x_i(g,q)^{-1}$ instead of $x_i(g,q)$.
Therefore $\sigma$ is a {\it circular permutation} of the four indices.

For $q<0$, the modular transform ${\tilde q}$ becomes complex, and we will make
no use of the above remarks. On the other hand, for $0< q <1$, the modular
parameter $\tau=it$ may be taken purely imaginary, and the modular transformation
reduces to $t\to 1/t$ hence ${\tilde q}$ is also real and in the range $(0,1)$. 
It is then easy to see that the modular transformation sends the solution $x_0(g,q)$
to $x_1(g,{\tilde q})$ and $x_1(g,q)$ to $x_2(g,{\tilde q})$ which leads to the
same $R_n$'s as $x_0(g,{\tilde q})$. In other words, the same physical solution
may be described by either some $q$ and the real determination $x_0(g,q)$ or
by its modular transform ${\tilde q}$ and the determination on the unit circle 
$x_1(g,{\tilde q})$. 

\subsec{Boundary condition at $n=L+1$}

We now implement the two-wall boundary condition described above, in which we
require $R_{L+1}=0$, or equivalently $u_{L+5}=0$.
This is achieved by demanding that $x^{L+6}= q^m$ for some $m\in \IZ$. From the 
above study, we have at our disposal $x$-solutions either real positive
and smaller than $1$,
or on the unit circle with positive imaginary part (we discard the equivalent $1/x$-solution).
This leaves us for {\it positive} $q$ with two possibilities: 
(i) $x=e^{2i\pi {k\over L+6}}$, $k=1,2,...$
corresponding to $m=0$ and (ii) $x=q^{m\over L+6}$, $m=1,2,...$
while at {\it negative} $q$, only the solution (i) survives. 
As the solution $R_n$ must be positive
by definition for $n=0,1,...,L$, it is easily verified that only $k=1$ in case 
(i) and $m=1$ in case (ii) are admissible to prevent sign changes for the $u_n$'s and the $R_n$'s.
In a more physical language, higher values of $k$ or $m$ correspond to 
higher modes with oscillations in the range $[0,L]$. Taking for instance 
$m=2$ corresponds to a first vanishing of $R_n$ at the coordinate $n=L/2-2$. 

For any given $L$, we may always pick the solution (i) and define $q_L(g)$ as the unique
real solution of 
\eqn\soli{x_1(g,q_L(g))=e^{2i\pi\over L+6}}
Introducing the notation $L(g)$ for the value of $L$ such 
that $x_1(g,q=0)=e^{2i\pi\over L+6}$, namely
\eqn\defgL{ L(g)={\pi\over \arctan\left((1-8g)^{1\over 4}\right)}-6}
we have that $q_L(g)>0$ for $L>L(g)$ and $q_L(g)<0$ for $L<L(g)$,
also valid for $g> 1/8$ with the convention that $L(g)=\infty$ in
this range.
The length $L(g)$ may be taken as a measure of
the typical extent in the
embedding space of the random trees at a fixed value of $g$, 
in the absence of walls. 

On physical grounds, we expect a qualitative change of behavior 
to occur at wall distances $L$ of the order of $L(g)$. For 
$L>>L(g)$, the tree behaves as a compact object of typical extent
$L(g)$ ``diffusing'' between the walls and feeling them only when
approaching  at distances of the order of $L(g)$ and smaller.
For $L<<L(g)$, the walls strongly squeeze the tree and $L$ is
the only relevant scale of the problem. In the first regime, 
the {\it profile} $\{R_n\}_{0\leq n\leq L}$, always maximal 
in the middle, is mainly constant and decreased significantly only at distances
of order $L(g)$ from the walls. In the second regime, the profile
will vary over the whole range between the walls. 
Alternatively, for a fixed value of $L$, we may invert these conditions into 
$q_L(g)>0$ for $g<g_L$ and $q_L(g)<0$ for $g>g_L$, where 
\eqn\defgLo{ g_L={1\over 8}\left(1-\tan^4\left({\pi\over L+6}\right)\right)} 
Note that $0<g_L<1/8$ for all $L=0,1,2,...$

Beside the solution \soli\ above, we have another
possibility for the choice of $q$ namely 
$q=q_L'(g)$ which solves the condition (ii) 
\eqn\solii{x_0(g,q'_L(g))=q'_L(g)^{1\over L+6}} 
This alternative solution exists for all $L>L(g)$ and is a priori distinct
from the previous solution \soli.
In practice, the {\it physical solution} for fixed $g$ and $L$ is {\it unique}
as one easily checks that
the solution \soli\ is the modular transform of \solii, namely
$q_L(g)={\tilde {q'}}_L(g)$ and $x_1(g,q_L(g))=x_0(g,q_L'(g))^{1\over \tau_L'(g)}$
with $q'_L(g)=e^{2i\pi \tau_L'(g)}$. The two choices \soli\ or \solii\ 
therefore lead to the same physical quantities $R_n$.

To obtain the combinatorial series expansions for the $R_n$'s in $g$ it is simpler
to work with the solution \solii. This is always possible as $g$ is
small (hence we may work in the regime $g<g_L$). Moreover, we have 
$x_0(g,q_L'(g))=g+O(g^2)$ while $q_L'(g)=g^{L+6}(1+O(g))$. This shows in particular
that the present $R_n$ and the former $R(g)$ of eq.\cata\ have the same 
expansion up to order Min$(n+1,L-n+1)$
in $g$, as expected. 

On the other hand, to have a more global approach it is best to work with 
the solution \soli, which is valid for any $g$. We may then
use the relations \gforeq\--\getRn\ with $\alpha=1/(L+6)$ 
to parametrize the solution with $q$ as follows 
\eqn\paraq{\eqalign{
g(q)&={\theta_1^4\left({1\over L+6}\right) \theta_1\left({4\over L+6}\right)\over 
\theta_1^5\left({2\over L+6}\right)}\cr
R(q)&= {\theta_1^3\left({2\over L+6}\right) \over \theta_1^2\left({1\over L+6}\right) 
\theta_1\left({4\over L+6}\right)} \cr
R_n(q)&= {\theta_1^3\left({2\over L+6}\right) \over \theta_1^2\left({1\over L+6}\right)
\theta_1\left({4\over L+6}\right)} \,{\theta_1\left({n+1\over L+6}\right)
\theta_1\left({n+5\over L+6}\right) \over\theta_1\left({n+2\over L+6}\right)
\theta_1\left({n+4\over L+6}\right)} \cr}}
with $\theta_1$ as in eq.\thetaone.
This form displays clearly the symmetry $R_n=R_{L-n}$ expected from the symmetry 
of the problem, as a consequence of the relation $\theta_1(1-z)=\theta_1(z)$.

\fig{Plot of $g(q)$ as given by eq.\paraq\ for $L=6$. The function decreases from
a maximum value $g_c(L)$ at some negative value $q_c(L)$ down to $0$ at $q=1$.
At $q=0$, we have $g=g_L$ as in eq.\defgLo. We have $g_L<{1\over 8}<g_c(L)$.}{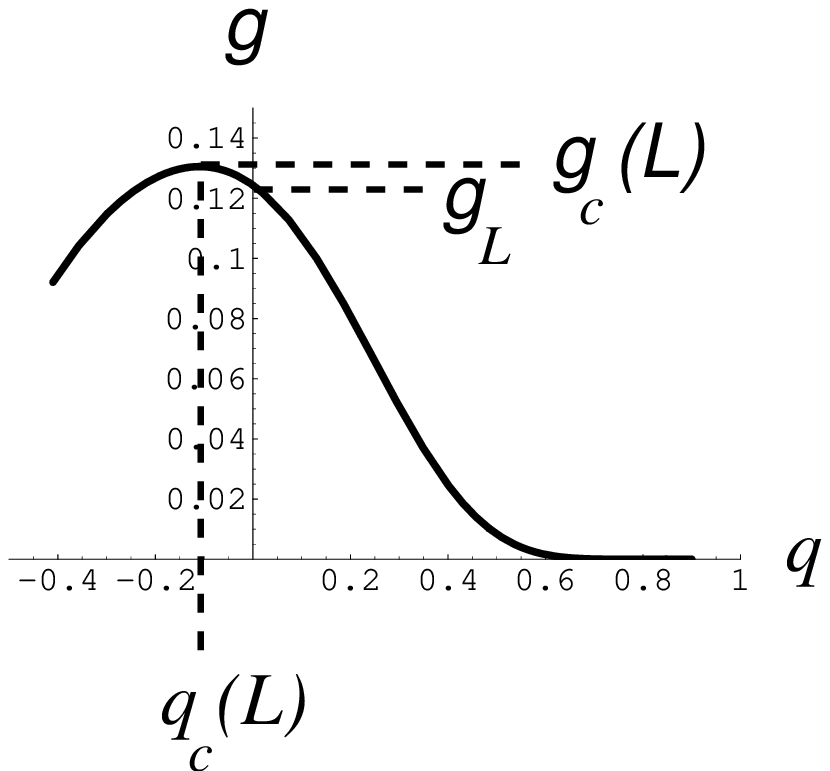}{7.cm}
\figlabel\gofq

For fixed $L$, the function $g(q)$ starts from $g=0$
at $q=1$ and increases as $q$ decreases, passing through $g_L$ of eq.\defgLo\ 
at $q=0$, and reaching a maximum
value $g_c(L)>1/8$ attained at some negative value of $q=q_c(L)$
(see Fig.\gofq\ for illustration). The quantity $g_c(L)$
is nothing but the radius of convergence of all the series in $g$ appearing in the
problem, and governs the leading
growth of the number of configurations as a function of the number $N$ of edges in the tree 
as $g_c(L)^{-N}$.
We have for instance $g_c(1)=1/4$, $g_c(2)=3-2\sqrt{2}$, $g_c(3)=4/27$. 
We also see that $g_c(L)\to 1/8$ when $L\to \infty$.
The branch of the solution for $q<q_c(L)$ is discarded as unphysical.

\fig{Typical profiles $\{R_n\}_{0\leq n\leq L}$ for $L=100$
as obtained from eq.\paraq\
for three particular values of $q$: (a) a positive value realizing $g\leq g_L$
where the profile is flat except for a region of extent $L(g)$ from the walls
(here $q=.25$ and $L(g)\simeq 18$) (b) $q=0$ ($g=g_L$) and (c) $q=q_c(L)$ ($g=g_c(L)$) where
the profile is slightly peaked in the middle.}{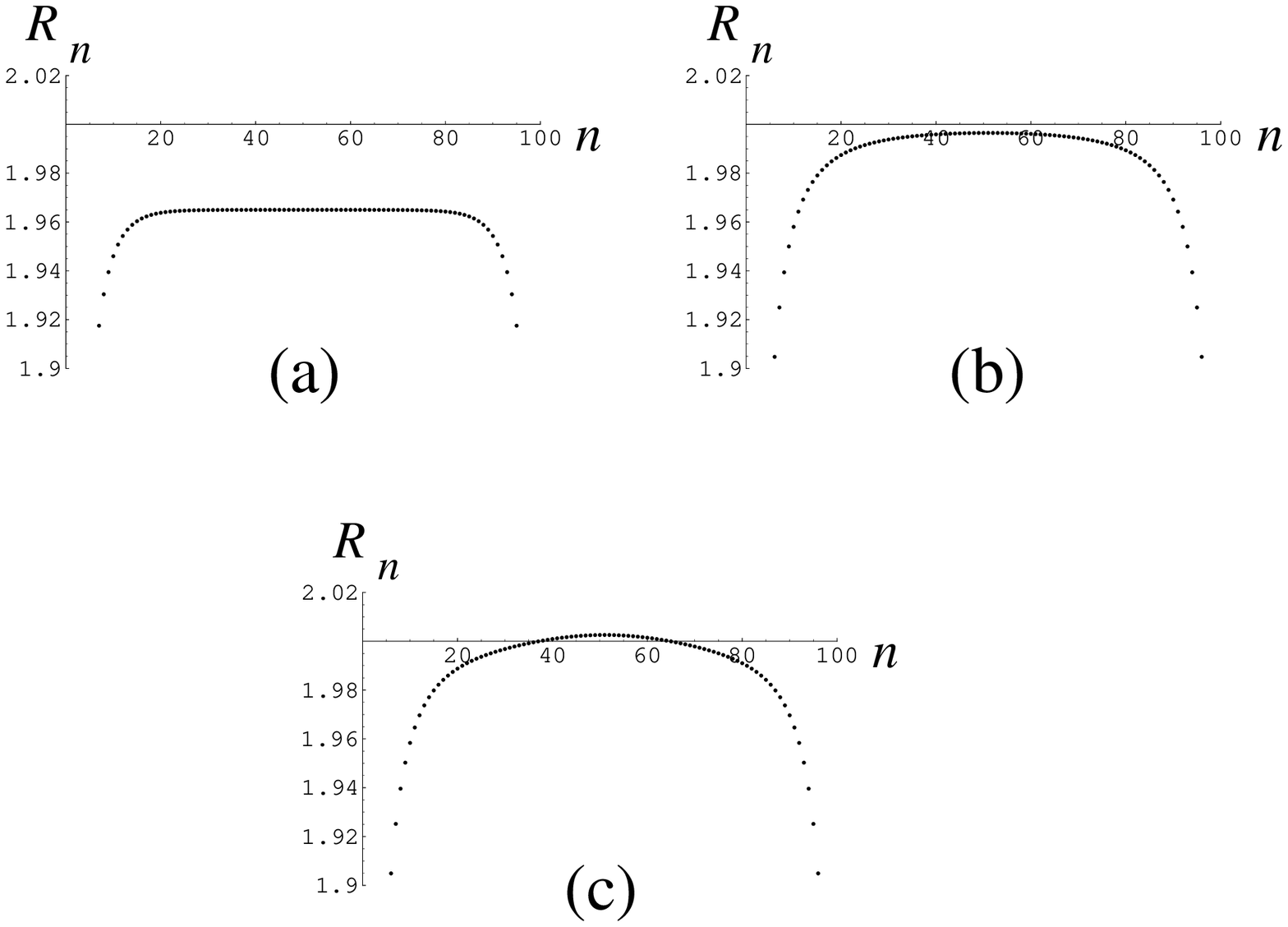}{14.cm}
\figlabel\profile

To conclude this section, let us use the solution \paraq\ to display for fixed $L$ the exact
profile $\{R_n\}_{0\leq n\leq L}$ for some particular values of $q$, namely
\item{}(a) a positive value of $q$ realizing $g<g_L$, in which case the profile is flat except
for a region distant by typically $L(g)$ from the walls 
\item{}(b) the value $q=0$ where $g=g_L$ and all theta functions degenerate into trigonometric
functions
\item{}(c) the negative value $q=q_c(L)$ corresponding to $g=g_c(L)$, where the profile
varies over the whole range $[0,L]$ and is slightly peaked in the middle
\par
These profiles are represented in Fig.\profile\ for $L=100$.

\newsec{Continuum limit}

For starters, let us derive the continuum limit of
the one-wall solution \onewsolu. It is reached by letting
$g \to 1/8$ and $n\ \to \infty$ simultaneously as:
\eqn\scag{g={1\over 8}(1-\epsilon^4),\qquad n={r\over \epsilon}}
with $\epsilon\to 0$ playing the role of the inverse of
a correlation length. This leads to
\eqn\scax{x(g)={1-\epsilon\over 1+\epsilon},\qquad R(g)={2\over 1+\epsilon^2}}
and finally, expanding $R_n$ up to order $2$ in $\epsilon$: 
\eqn\scaRn{R_n=2(1-\epsilon^2 {\cal U}(r))}
where ${\cal U}$ is the scaling function
\eqn\Usca{{\cal U}(r)=1+{3\over \sinh^2(r)}} 
describing the ``repulsive potential" felt by the tree as a function of the 
rescaled distance $r$ from the wall at $r=0$.

In the presence of the second wall, we let in addition $L\to \infty$
by keeping the quantity $\omega=(L+6)\epsilon/2$ fixed.
This ratio of the two characteristic lengths of the problem, namely
$L$ and the typical extent of the unconstrained tree $L(g)\sim\pi/\epsilon$ is the 
only physical scale surviving the continuum limit.
Expanding the rhs of the first line of equation \paraq\ at small $\epsilon$
up to order $4$, we get
\eqn\gexpan{g={1\over 8} \left(1-{\epsilon^4 \over \omega^4}\left(
{5\over 32}\left({\theta_1'''(0)\over \theta_1'(0)}\right)^2-{3\over 32}{\theta_
1^{(5)}(0)
\over \theta_1'(0)} \right)\right)}
to be identified with $g=1/8(1-\epsilon^4)$ as in eq.\scag. This fixes
implicitly the value of $\tau$ as a function of $\omega$ by
\eqn\fixom{ \omega^4={5\over 32}\left({\theta_1'''(0)\over \theta_1'(0)}\right)^
2-{3\over
32}{\theta_1^{(5)}(0)
\over \theta_1'(0)} }
We may now expand $R$ and $R_n$ as given by eq.\paraq\ up to order $2$ 
in $\epsilon$ to obtain the relevant scaling function
\eqn\tendto{\eqalign{
R&= {2}\left(1- {\epsilon^2\over 4\omega^2}{\theta_1'''(0)\over \theta_1'(0)}
\right)\cr 
R_n&= 2\left(1-\epsilon^2 {\cal U}(r)\right)\cr}}
with $r$ as in eq.\scag\ and where the scaling function ${\cal U}(r)$ is
now given by
\eqn\alU{{\cal U}(r)={\theta_1'''(0)\over 4\omega^2 \theta_1'(0)}
-3{d^2\over dr^2}{\rm Log}\, \theta_1\left({r\over 2 \omega}\right) 
= 3\wp (r)}
where we have identified the Weierstrass $\wp$ function \BAT\ with half-periods
$\omega$ and $\omega'=\tau \omega$ where $\tau$ is fixed by eq.\fixom,
which amounts to
\eqn\gdeux{g_2(\omega,\omega')={4\over 3}}
where $g_2$ is the first elliptic invariant of the Weierstrass function.
The scaling function ${\cal U}(r)$ may be viewed as the potential felt
by the random tree in the presence of the two walls.

When sending the second wall to infinity, i.e. taking $\omega\to \infty$,
we immediately recover the above result \Usca\ as $g_2=4/3$ fixes
the second half-period $\omega'=i \pi/2$, in which case the scaling 
function degenerates into \Usca.
Taking $L=L(g)$, i.e. $\omega=\pi/2$, fixes $\omega'=i \infty$ 
which corresponds to $q=0$, in which case the scaling 
function degenerates into a trigonometric function
\eqn\trigop{ {\cal U}(r)={3\over \sin^2(r)}-1}
The values of $\omega\in [\pi/2,\infty]$ ($L\geq L(g)$) correspond to taking $q\in [0,1]$,
while those in $[0,\pi/2]$ ($L\leq L(g)$) are obtained for $q\in [-q_*,0]$, where
$q_*=-\lim_{L\to \infty} q_c(L)=e^{-\pi \sqrt{3}}$.

Note that we could have derived the scaling limit of $R_n$ without
using the explicit solution \paraq\ by plugging the ansatz 
$R_n=2(1-\epsilon^2 {\cal U}(r))$ in the original equation
\recu\ and expanding up to order $4$ in $\epsilon$ to obtain a 
differential equation for ${\cal U}$, namely
\eqn\prewei{{\cal U}''=2({\cal U}^2-1) }
and require that ${\cal U}(r)$ diverge at $r=0$ and $r=2\omega$, with
no divergence in-between. This equation is to be compared with
that satisfied by $\wp$, namely $\wp''=6\wp^2-g_2/2$, which allows
to identify ${\cal U}(r)=3\wp(r)$ provided $g_2=4/3$. 

\newsec{A simple application: escape probability from a fixed domain for a spreading population}

As mentioned in the introduction, rooted planar trees may be used to
model discrete branching evolution processes.
Let us assume for instance that an initial parent
individual (materialized
by the root vertex) gives rise in one generation to a number $k\geq 0$ of 
children individuals with probability $p_k$,
each child itself independently giving rise to subsequent generations with the same
probabilities. 
The resulting genealogical tree is nothing but a {\it planar} tree
(without labels).
In the following, we concentrate on the most natural choice 
$p_k=(1-p)p^k$, with $0\leq p\leq 1$, and where the prefactor $(1-p)$
ensures the correct normalization and may be interpreted as the probability 
of death without descendents. This choice is expected to capture all possible 
physics of the problem, as the parameter $p$ allows to explore all possible values of the 
average number of children $p/(1-p)$, known to be the only relevant quantity
in the problem \GALWA.

We may now turn the branching process into a {\it spatial} branching process 
by allowing the individuals to spread in a one-dimensional
target space.
More precisely, we consider a discrete version of the problem
in which the individuals may occupy integer-valued positions, 
with the diffusion rule that 
each child lives at a position differing by $\pm 1$
from that of its parent, with an equal probability $1/2$. 
Using these positions as vertex labels, 
we may write the following master equation for the extinction probability
of a family 
\eqn\death{ E_n(T)= {1-p\over 1-{p\over 2}(E_{n+1}(T-1)+E_{n-1}(T-1))} }
where $E_n(T)$ stands for the probability that an individual sitting
at position $n$ has no more descendent at generation $T$,
with the initial condition $E_n(0)=0$.
Eq.\death\ is obtained by enumerating all possible 
configurations of the first generation children, and noting that the joint
probability of extinction of all their descendents
is the product of individual extinction probabilities before generation $T-1$.  
Comparing eq.\death\ to eq.\recu, we see that 
$E_n\equiv \lim_{T\to \infty} E_n(T)$
obeys the same equation \recu\ as $(1-p)R_n$, provided we set
\eqn\gset{ g={p(1-p)\over 2} }
This allows to identify $E_n=(1-p)R_n$ by noting that 
this choice corresponds precisely to the stable fixed point of the recursion relation \death.
We may therefore interpret $(1-p)R_n$ as the probability of extinction of a family
whose first generation's parent sits at position $n$.
Accordingly, we interpret $1-(1-p)R_n$ as the survival probability for such a family.
 
For unconstrained positions (no wall in the former language), we simply have
a translation invariant probability of survival 
\eqn\simsn{  S(p)\equiv {2p-1+|2p-1| \over 2p} =\left\{ \matrix{ 0 & p\in [0,{1\over 2}]\cr
{2p-1\over p} & p\in [{1\over 2},1]\cr} \right. }
obtained by substituting $g=p(1-p)/2$ into eq.\cata. This result is totally insensitive to the
diffusion process, and is the same as that for unlabeled trees. It displays a
first order singularity (discontinuous derivative) at $p=p_c={1\over 2}$. 
This classical result in the theory of branching processes \GALWA\ is a
particular case of a more general statement for so-called Galton-Watson
processes that the genealogical tree is almost surely finite 
(here $S(p)=0$) if and only if 
the average number of children is less or equal to one (here
corresponding to $\sum j p_j=p/(1-p)\leq 1$, namely $p\leq 1/2$). 

Apart from the extinction probability $E_n(T)$,
one natural quantity to study is the probability $C_n(T)$ for the population spreading from a position
$n$ to remain {\it confined} within a given connected domain $\cal D$ until 
generation $T$. 
This quantity is readily seen to obey the same equation \death\ as $E_n(T)$
for all $n$ in ${\cal D}$, but with a different initial
value at $T=0$, namely $C_n(0)=1$, and also the condition that
$C_n(T)=0$ for all $n$ outside of ${\cal D}$.
We distinguish the two possible cases where (i) $\cal D$ is 
a half-line, say $[0,\infty)$ or (ii) $\cal D$ is a segment, say $[0,L]$.
The only relevant conditions are at the boundary of the domain and 
read respectively (i) $C_{-1}(T)=0$ and (ii) $C_{-1}(T)=C_{L+1}(T)=0$.
In the limit when $T\to \infty$, we may therefore identify 
$C_n(\infty)=(1-p)R_n$ with $R_n$ given by our (i) one-wall and (ii) 
two-wall solutions of Sects. 2 and 3.
The results will be best expressed in terms of the quantity
$S_n = 1 - C_n(T = \infty)$
which is nothing but the probability for the population to escape from the domain.

In the one-wall case,
the solution \onewsolu\ leads to the population's escape probability 
\eqn\surone{S_n= 1 - {1-|2p-1|\over 2p} {(1-x^{n+1})(1-x^{n+5})\over (1-x^{n+2})(1-x^{n+4})} }
where
\eqn\valux{ x=x(p)\equiv {1-|2p-1|^{1\over 2} \over 1+|2p-1|^{1\over 2}} } 
Note that for any fixed $n\geq 0$ the escape probability $S_n$ is strictly positive
as soon as $p>0$, and moreover that it still displays a singularity at $p=p_c=1/2$
but of weaker {\it third order} 
type (discontinuity of the third derivative) as is readily seen by expanding $S_n$
up to order $3$ in powers of $|2p-1|$.
Note also that $\lim_{n\to \infty} S_n=S(p)$ as in eq.\simsn, expressing the equivalence 
in probability
between surviving forever and reaching infinitely distant points.
Note finally the following simple
expression for the escape probability $S_n$ at the transition point $p={1\over 2}$:
\eqn\transis{ S_n(p={1\over 2})= {3 \over (n+2)(n+4)}, \quad n\geq 0 }
The scaling limit \scag\ may be used to study the vicinity of the transition point
by setting $p=p_c(1+\eta \epsilon^2)$ with $\eta=\pm 1$ according to whether we approach
the transition from above or below. Eq.\scaRn\ allows to interpret
the scaling function ${\cal U}(r)$ as describing the scaling behavior
of the escape probability $S_n\sim {\cal S}_n$ around $p={1\over 2}$, with 
\eqn\sursca{ {\cal S}_n= \epsilon^2 ({\cal U}(n\epsilon)+\eta)=
|2p-1|\left({3\over \sinh^2(n|2p-1|^{1/2})} +1\right) +(2p-1)}
valid in the scaling region of large $n$ and $n\epsilon=O(1)$.
This scaling function displays clearly the above-mentioned third order transition
with
\eqn\devepsi{ {\cal S}_n={3\over n^2}+ (2p-1) + {n^2\over 5} (2p-1)^2-{2n^4\over
63}|2p-1|^3+O((2p-1)^4n^6)}

\fig{Escape probabilities $S_n$ of eq.\surone\ as functions of $p$ in the
presence of one wall,
(a) for $p\in [0,1]$ and from top to bottom $n=0,1,2,3,4$ as well as 
$n=\infty$ in which case we recover the no-wall solution $S(p)$ of eq.\simsn;
(b) for $p$ in the critical region $p\sim {1\over 2}$ and from top to bottom
$n=5,7,12$, together with the expected scaling limits ${\cal S}_n$ as given by eq.\sursca\
with a proper shift of $n\to n+3$ ensuring the perfect matching of the curves.}{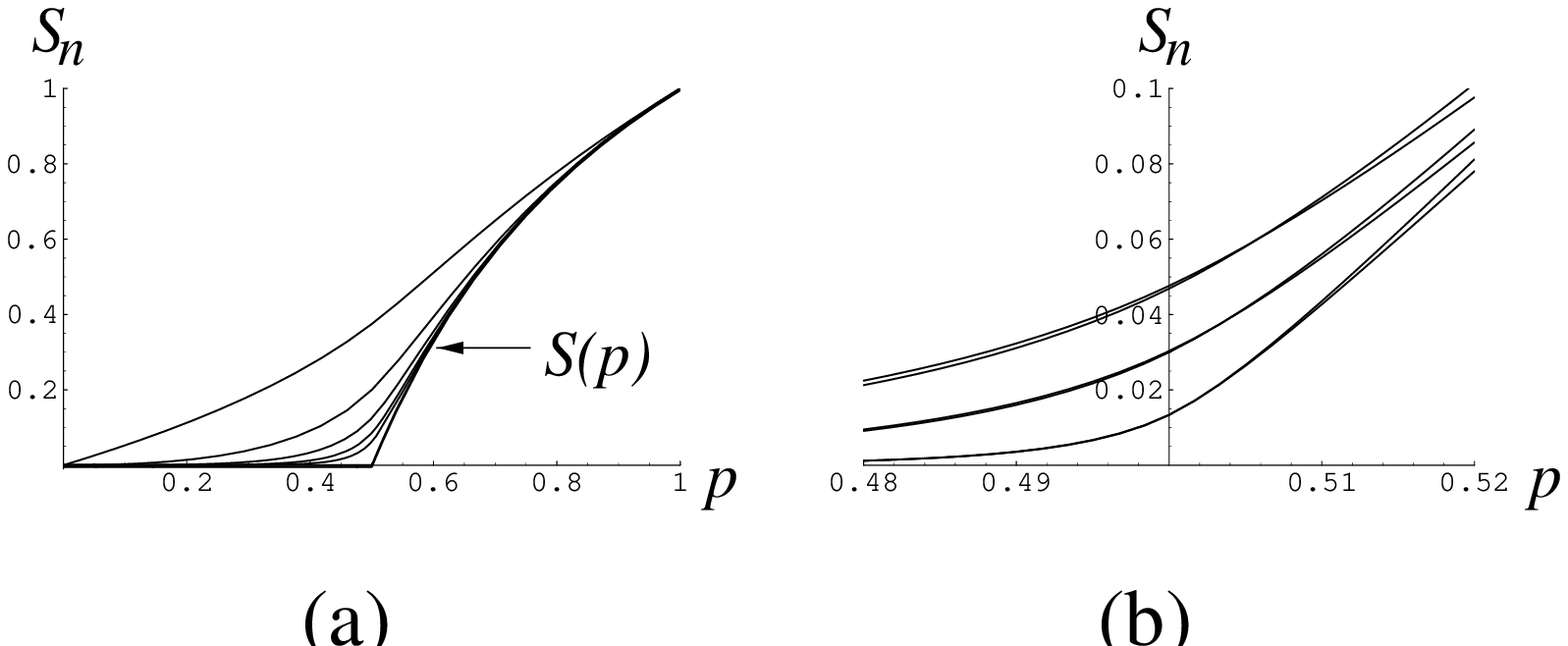}{14.cm}
\figlabel\survonewall
We have represented in Fig.\survonewall\ the escape probabilities $S_n(p)$ and their
limit $S(p)$
as functions of $p\in [0,1]$.
We have also blown out the critical region around $p=1/2$ to compare the exact solution
with its scaling limit.

In the case of two walls, 
the solution \paraq\ leads directly to the probability $S_n=1-(1-p)R_n$
of escaping from the interval $[0,L]$. 
\fig{Plot of the increase in the escape probability 
$S_n^{(L)}(p)-S_n(p)$ from the one-wall situation to that with two walls
for $L=5$ and $n=0,1,2$ (bottom to top).}{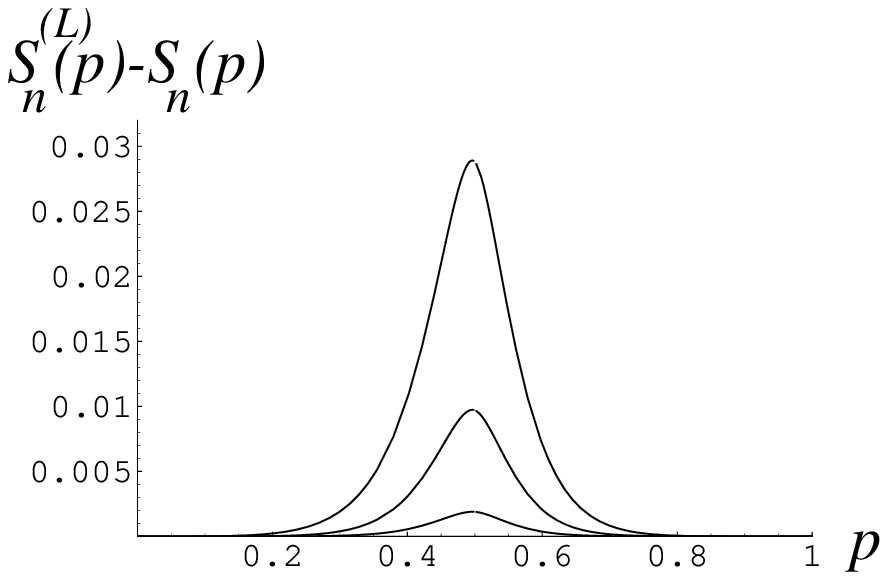}{9.cm}
\figlabel\diffsur
For illustration, 
we have displayed in Fig.\diffsur\ the increase in the escape probability 
$S_n^{(L)}(p)-S_n(p)$ from the one-wall situation to that with two walls
for $L=5$ and $n=0,1,2$, as functions of $p\in [0,1]$. The increase is maximal
at $p={1\over 2}$.

For finite $L$, the critical value $g_c(L)>1/8$ is never attained 
as $g=p(1-p)/2\in [0, 1/8]$ for $p\in [0,1]$, hence the singularity at 
$p={1\over 2}$ is suppressed. 
However it is restored in the scaling limit where $L\to \infty$ with 
$L|2p-1|^{1/2}$ fixed, as
$g_c(L)\to 1/8$ in this case. The scaling function
${\cal U}(r)$ given by eq.\alU\ again describes the scaling behavior
of the escape probability $S_n\sim {\cal S}_n$ in the vicinity of $p={1\over 2}$, with the
result: 
\eqn\surscab{ {\cal S}_n= 3 |2p-1| \wp(n|2p-1|^{1/2})+(2p-1) }
where the Weierstrass $\wp$ function must be taken with fixed half-periods
$\omega=L |2p-1|^{1/2}/2$ and $\omega'=\tau \omega$, such that
$g_2(\omega,\omega')={4\over 3}$. We again note that ${\cal S}_n$ displays a third
order singularity at $p={1\over 2}$ by expanding 
\eqn\devewp{{\cal S}_n={3\over n^2}+ (2p-1) + {3\over 20} g_2(\omega,\omega') n^2 (2p-1)^2+{3\over 28}
g_3(\omega,\omega') n^4 |2p-1|^3+O((2p-1)^4n^6)}
where $g_2$ and $g_3$ stand for the elliptic invariants of the Weierstrass function
and with the constraint that $g_2(\omega,\omega')=4/3$ which fixes $\omega'$ and
consequently $g_3$ as functions of $\omega=L|2p-1|^{1/2}/2$.
Note that the singularity of ${\cal S}_n$ at $p=1/2$ disappears at the modular 
invariant point $\tau=i$ i.e. $q=e^{-2\pi}$, where 
$g_3(\omega,i\omega)=0$, which causes all odd powers of $|2p-1|$ to vanish in the series 
expansion \devewp.

Of particular simplicity is the case
when $q=0$ in eq.\paraq, corresponding to $g=g_L$ as in eq.\defgLo, i.e. 
$p=(1-\tan^2(\pi/(L+6)))/2$ or $p=(1+\tan^2(\pi/(L+6)))/2$.
The formula for the associated escape probability $S_n$ reads
\eqn\soltrigo{ S_n= \left\{ \matrix{ {1\over \cos\left({2\pi \over L+6}\right)}
\left(
{\sin\left({\pi\over L+6}\right) \sin\left({3\pi\over L+6}\right)\over
\sin\left(\pi {n+2\over L+6}\right) \sin\left(\pi {n+4\over L+6}\right)} 
-2 \sin^2\left({\pi\over L+6}\right)\right) & 
p={1\over 2}\left(1-\tan^2\left({\pi\over L+6}\right)\right) \cr
{\sin\left({\pi\over L+6}\right) \sin\left({3\pi\over L+6}\right)\over 
\sin\left(\pi {n+2\over L+6}\right) \sin\left(\pi {n+4\over L+6}\right)} & p=
{1\over 2}\left(1+\tan^2\left({\pi\over L+6}\right)\right) \cr}\right. }
This allows for framing the exact value of $S_n$ at $p=1/2$ between these
two values for all $L$. For large $L$, we may identify $\omega=\lim_{L\to \infty}
L |2p-1|^{1/2}/2=\pi/2$ for both values of $p=(1\pm\tan^2(\pi/(L+6)))/2$, 
and $\tau=i \infty$ to ensure $q=0$, in which case
the scaling function reduces to \trigop\ and
therefore eq.\surscab\ turns into
\eqn\therS{ {\cal S}_n=|2p-1|\left({3\over \sin^2(n|2p-1|^{1/2})} -1\right) +(2p-1)}  
in agreement with the large $L$ limit of eq.\soltrigo.

\newsec{Another solvable case of tree embedding: dilute SOS model on a random tree}

We may consider a slightly different version of labeled trees, 
in which we impose the weaker constraint that any two adjacent 
vertices of the tree must have labels differing by $\pm 1$ {\it or} $0$. 
As shown in Ref.\CS,
this version of labeled trees is that involved in the enumeration of 
tetravalent planar graphs.  
In the language of spreading of a population, a child may now stay at 
the same position as its parent. This corresponds to a 
dilute SOS version of the case studied in this paper.

The main recursion relation is now replaced by
\eqn\quarecu{ R_n={1\over 1-g(R_{n+1}+R_n+R_{n-1})} }
where $R_n$ is the generating function for rooted trees with root
vertex labeled by $n\in \IZ$, and a weight $g$ per edge.
Again, we may consider three types of boundaries: no wall, 
one wall and two walls.
The no-wall case is easily solved, with all the $R_n$'s equal
to the solution of $R=1/(1-3gR)$ with $R=1+O(g)$, namely 
\eqn\zerosol{ R=R(g) \equiv {1-\sqrt{1-12 g}\over 6g} }
The one-wall case corresponds to setting $R_{-1}=0$
and only considering the $R_n$'s for $n\geq 0$. It was solved in Ref.\GEOD, with
the result
\eqn\resuone{
R_n=R {u_n u_{n+3}\over u_{n+1} u_{n+2} }, \qquad u_n=x^{n+1\over 2}-x^{-{n+1\over 2}} }
with $R=R(g)$ of eq.\zerosol, and
where $x$ is the solution of $x+1/x+4=1/(g R)$ with, say, modulus less than $1$.
Note the slight difference in the index shifts when compared with eq. \onewsolu.
The main recursion relation reduces this time to a quartic
equation for the $u_n$'s:
\eqn\qartiu{
u_nu_{n+1}u_{n+2}u_{n+3}= {1\over R} u_{n+1}^2u_{n+2}^2+g R(u_{n-1}u_{n+2}^2u_{n+3}+
u_n^2u_{n+3}^2+u_n u_{n+1}^2u_{n+4})}
supplemented by the initial condition $u_{-1}=0$.
Eq.\qartiu\ is
easily checked for all $k$ by setting $1/R=(x^2+x+1)/(x^2+4x+1)$ and $gR=x/(x^2+4x+1)$.  
Finally, in the two-wall case where we require $R_{-1}=R_{L+1}=0$ and only consider
$n=0,1,2,...,L$, we have found the elliptic solution
\eqn\eliqua{\eqalign{
R_n&=R {u_n u_{n+3}\over u_{n+1} u_{n+2} }\cr 
u_n&=\theta_1((n+1)\alpha) \cr}}
with $x=e^{2i\pi \alpha}$ and $\theta_1$ as in eq.\thetaone, and
which solves eq.\qartiu\ provided we take
\eqn\wetak{\eqalign{ R&= 4 {\theta_1(\alpha) \theta_1(2\alpha)\over
\theta_1'(0)\theta_1(3\alpha)} \left({\theta_1'(\alpha)\over \theta_1(\alpha)}-{1\over 2}
{\theta_1'(2\alpha)\over \theta_1(2\alpha)}\right)\cr
g&={\theta_1'(0)^2\theta_1(3\alpha)\over 16\theta_1(\alpha)^2\theta_1(2\alpha) 
\left({\theta_1'(\alpha)\over \theta_1(\alpha)}-{1\over 2}
{\theta_1'(2\alpha)\over \theta_1(2\alpha)}\right)^2}\cr} }
The boundary conditions are again satisfied for two choices of the parameter $x$:
(i) $x=e^{2i\pi/(L+5)}$ and (ii) $x=q^{1/(L+5)}$ (when $q\geq 0$), the latter 
leading to the same physical solution as the former by modular invariance. 
Picking again the first solution, we must take $\alpha=1/(L+5)$, and may view
the equations for the solution as parametrized by $q$.
This leads to the continuum limit, upon taking
$g=(1-\epsilon^4)/12$, $n=r/\epsilon$, 
$2\omega= (L+5)\epsilon$ and $R_n=2(1-\epsilon^2{\cal U}(r))$.
We end up
with a scaling function ${\cal U}(r)=2 \wp(r)$ in terms of the Weierstrass $\wp$ function
with half-periods $\omega$ as above and $\omega'$ fixed by now requiring that
$g_2(\omega,\omega')=3$.  
We also recover the one-wall case by taking the limit $L\to\infty$, namely $\omega=\infty$
while $\omega'=i\pi/\sqrt{6}$, in which case ${\cal U}(r)=1+3/\sinh^2(\sqrt{3/2} r)$,
a result already obtained in Ref.\GEOD.

Note finally that all scaling functions coincide with those of Sect.4 up to 
a global rescaling $r\to \sqrt{3\over 2} r$ and $\omega\to \sqrt{3\over 2} \omega$.
This confirms the expected universality of the continuum limit.

\newsec{Conclusion}

In this paper, we have extensively studied a model of random rooted planar 
trees embedded in a discrete one-dimensional target space. In particular, we have derived
explicit expressions for the partition function of the model with various
target spaces, namely the whole integer line, a half-line, and a segment. 
To obtain these, we have
shown that the partition functions actually obey recursion relations and that the
particular target at hand translates into various boundary conditions.  
We have also derived the corresponding scaling functions in the continuum
limit for which the recursion relations turn into differential
equations. 
A different approach, popular among probabilists, consists in studying 
directly the continuum limit of embedded random trees in the form
of continuum spatial branching processes \LEGAL, giving rise
to partial differential equations. It should be possible in this 
context to solve these equations with wall-type boundary conditions,
and to recover our continuum results. Some work in this direction
appeared recently \DELMAS, where an analogue of our one-wall case can be found.

The striking simplicity of the solutions \onewsolu\-\twowsolu\ as well as 
\resuone\-\eliqua\ are directly linked to the ``integrability" of the corresponding
non-linear recursion relations \recu\ and \quarecu\ respectively. 
One possible explanation for the integrability uses the interpretation of 
labeled trees in the context of planar graph enumeration.
More precisely, as shown in detail in appendix A below, the SOS and dilute SOS
models on trees respectively occur in the enumeration of rooted planar
Eulerian triangulations (i.e. triangulations with bicolored faces) and of
rooted planar quadrangulations. In both cases, the labels $n$ correspond
to geodesic distances along the graph from the root. 
This reformulation suggests a possible connection with matrix models,
known to be integrable. 
As already hinted 
in Ref.\GEOD\ in the context of graph enumeration, the equations \recu\
and \quarecu\ 
are very similar to those arising in the context of the
matrix models used for generating Eulerian triangulations and quadrangulations
respectively, namely $R_n=(n/N)/(1-g(R_{n+1}+R_{n-1}))$ 
and $R_n=(n/N)/(1-g(R_{n+1}+R_n+R_{n-1}))$, 
where $N$ is the size of the matrices.  
The remarkable point is that the index $n$ is no longer related to the geodesic 
distance along the graphs, but to their {\it genus}. 
Soliton theory seems to indicate that
integrability survives when changing the recursion into 
$R_n=(\alpha+\beta n)/(1-g(R_{n+1}+R_{n-1}))$ or
$R_n=(\alpha+\beta n)/(1-g(R_{n+1}+R_n+R_{n-1}))$ for some constants
$\alpha$ and $\beta$, which could interpolate between the two problems.

Beyond the two (quadratic) examples of this paper, we have at our disposal a host
of non-linear recursion relations all used for enumerating possibly decorated
planar graphs while keeping track of some geodesic distances, and which were
found to be integrable as well (see Ref.\GEOD\ for details). The solutions
display some multicritical behavior corresponding to higher order critical points,
with non-trivial hierarchies of scaling functions. It would be interesting to
find some proper interpretation of those equations in the context of embedded
trees or alternatively of population-spreading processes.

It would also be interesting to classify the target spaces leading
to integrable models of embedded trees, or alternatively to spot among all possible discrete
spatial branching processes those with integrability properties. We may then hope,
by changing the nature of the discrete target space, to be able to
reach new critical points.

\appendix{A}{Graph interpretation}

As mentioned above, well-labeled trees, i.e. trees with non-negative labels
corresponding to the one-wall situation in the dilute SOS version of
Sect.6 were introduced in Ref.\CS\ in the context of graph enumeration.
More precisely, it was shown that there exists a bijection between these well-labeled
rooted planar trees and rooted planar quadrangulations. This allows to
interpret the quantity $R_n$ of eq.\resuone\ as the generating function for planar quadrangulations
with both a marked (origin) vertex and a marked oriented edge linking a vertex
with geodesic distance $m$ from the origin to a vertex with geodesic distance $m+1$
from the origin with $m\leq n$, and with an activity $g$ per vertex. 
Beside this equivalence, there exists yet another bijection, now in the dual language,
between rooted planar tetravalent graphs (dual to the above quadrangulations)
and decorated (so-called blossom-) binary trees \SCH. 
This bijection was extended so as to keep track of geodesic distances
between faces in Ref.\GEOD. In this language, the generating function $Z_n$
for two-leg planar tetravalent graphs with the two legs
distant by at most $n$ was shown to obey the recursion relation 
$Z_n=1+g Z_n(Z_{n+1}+Z_n+Z_{n-1})$, with $Z_{-1}=0$, a direct consequence
of the above bijection with blossom binary trees. This equation is nothing but 
yet another form of eq.\quarecu\ and allows to identify $Z_n=R_n$ of eq.\resuone.

Remarkably, our slightly simpler equation \recu\ also
admits two analogous interpretations in terms of graphs,
now related to the enumeration of rooted planar Eulerian triangulations (i.e. triangulations
with bi-colored faces, say in black and white) or dually to that of rooted trivalent 
bipartite planar graphs (say with black and white vertices). 
In this dual language, we may rely
on a bijection \BMS\ between rooted trivalent
bipartite planar graphs 
and properly decorated binary trees. Keeping track of the graph-geodesic distances
in these binary trees leads to the equation $Z_n=1+g Z_n(Z_{n+1}+Z_{n-1})$ 
for the generating function $Z_n$ of two-leg trivalent bi-colored
planar graphs with the two legs attached to vertices of opposite colors and at
geodesic distance at most $n$. In this approach,
the proper definition of the geodesic distance on the graphs 
makes use of oriented paths linking faces, with the constraint that a step across 
an edge between two faces always leaves the black vertex on the left. 
The equation for $Z_n$ is yet another form of eq.\recu\ which allows to identify $Z_n=R_n$ of
eq.\onewsolu. 

\fig{A sample Eulerian triangulation (a) with a marked origin vertex
(labeled $0$) and a marked oriented edge (big empty arrow). We have indicated
for each vertex its geodesic distance from the origin (respecting the edge-orientations).
A labeled rooted tree (b) is obtained by retaining for each clockwise-oriented
triangle the edge connecting the two farthest vertices from the origin, and
picking as the root the end of the previously marked edge. The vertex labels on the tree
are simply the distances of the graph vertices from the origin minus one.}{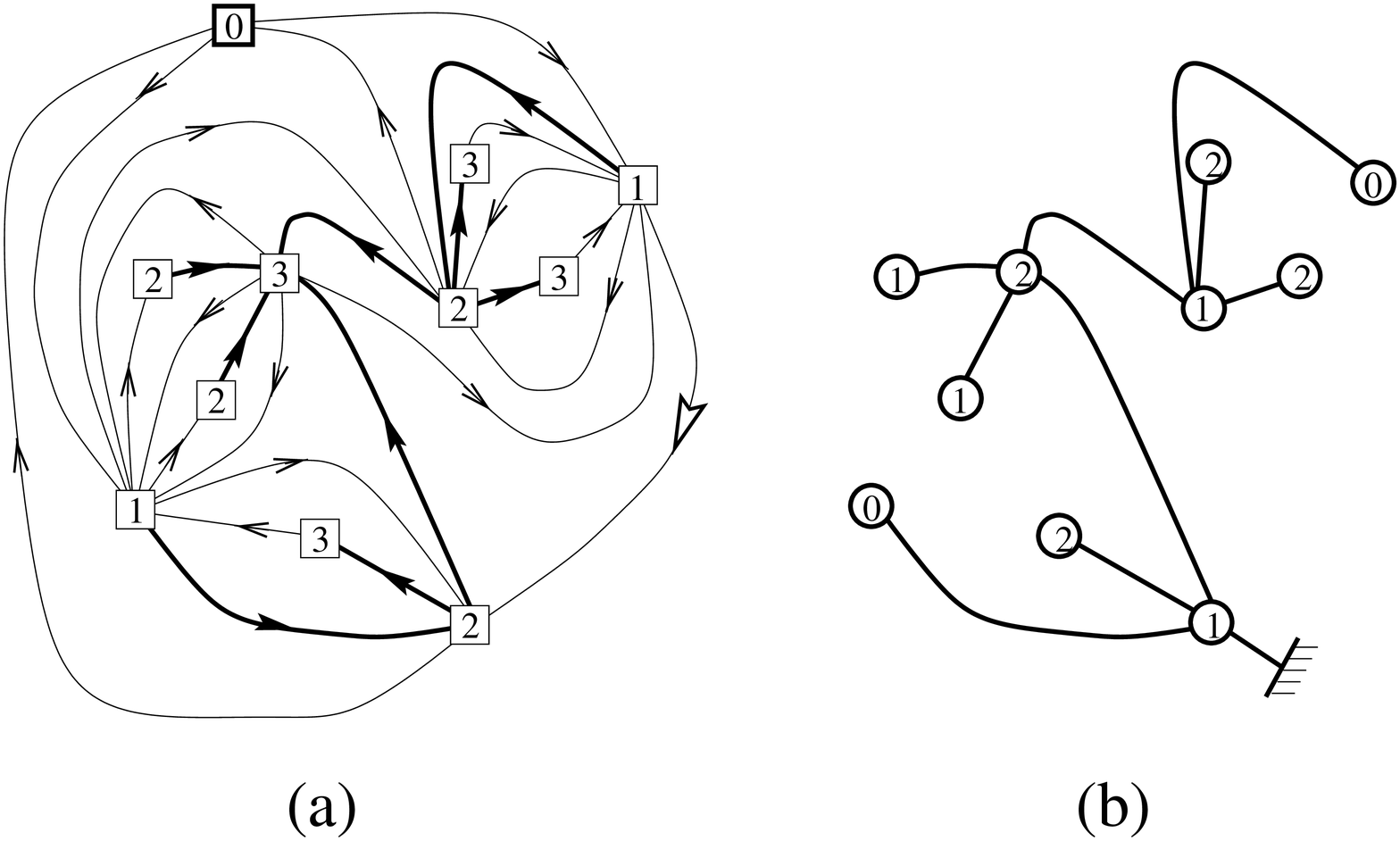}{13.cm}
\figlabel\euler

In terms of triangulations, a bijection similar to that of Ref.\CS\ may be established
between rooted planar Eulerian triangulations and the well-labeled (SOS) trees
corresponding to the one-wall situation of Sect.2  as follows. First we replace
the face-bicoloration by the compatible orientation of all edges in such a way
that each triangle is either clockwise- or counterclockwise oriented. This allows to define
the geodesic distance from a vertex to another as the length of a minimal path respecting
the orientation of edges. 
Picking as origin some vertex,
a well-labeled tree is obtained by retaining for
each clockwise-oriented triangle the edge linking the two farthest vertices from    
the origin, and labeling each vertex by its distance from the origin minus one,
as illustrated in Fig.\euler\ (see Ref.\LERETOUR\ for details and proofs). 
This allows to interpret $R_n$ as the generating function for planar Eulerian triangulations
with a marked (origin) vertex 
and a marked oriented edge linking a vertex
with geodesic distance $m$ from the origin to a vertex with geodesic distance $m+1$
from the origin with $m\leq n$, and with an activity $g$ per vertex.

In the language of graphs, we may use our two-wall solutions to enumerate 
{\it bounded} graphs as follows.  The quantity 
\eqn\gengj{ G_n^{(L)}=R_n^{(L)} -R_{n-1}^{(L-1)} }
where $R_n^{(L)}$ is the solution of eq.\recu\ (resp. \quarecu) with two walls at positions
$-1$ and $L+1$, is the generating function for Eulerian
triangulations (resp. for quadrangulations) with a marked (origin) vertex,
a marked oriented edge linking a vertex
with geodesic distance $n$ from the origin to a vertex with geodesic distance $n+1$
from the origin and which are ``bounded" in the sense that the geodesic distance of 
{\it all the vertices} from the origin is less or equal to $L+1$.
As an example, the rooted Eulerian triangulation of Fig.\euler\ contributes
to $G_{1}^{(L)}$ for all $L\geq 2$.

\listrefs
\end